\documentclass{iopart}
\usepackage{graphicx}
\usepackage{iopams}
\usepackage{amssymb}

\newcommand\UPDATEBIB[1]{}
\newcommand\UPDATEBIBNO[1]{#1}

\newcommand{\cut}[1]{}

\newcommand{\INPAPER}[1]{}

\newcommand{\<}{\left\langle}
\renewcommand{\>}{\right\rangle}
\def\rmd{d}

\newcommand{\mQ}{{\cal Q}}

\def\se{{\tt se}}
\def\SNRb{{\mathrm{SNR_b}}}
\def\BER{{\mathrm{BER}}}

\def\Extr{{\mathrm{Extr}}}

\newcommand{\sign}{{\mathrm{sign}}}

\def\be{\begin{equation}}
\def\ee{\end{equation}}

\def\vy{{\mbox{\boldmath{$y$}}}}

\def\vH{{\mbox{\boldmath{$H$}}}}

\def\vb{{\mbox{\boldmath{$b$}}}}
\def\vtau{\mbox{\boldmath{$\tau$}}}

\def\vs{\mbox{\boldmath{$s$}}}
\def\sk{{\vs_k}}
\def\vomega{{\mbox{\boldmath{$\omega$}}}}

\def\vones{\mbox{\boldmath{$1$}}}

\def\vtau{{\mathbf{\tau}}}

\def\ms{{\mathbb S}}
\def\Ms{{\mathbb S}}

\newcommand{\orderedL}[1]{{\langle {#1}_1, \ldots, {#1}_{L} \rangle}}

\begin{document}

\title{Composite CDMA - A statistical mechanics analysis}
\author{Jack Raymond and David Saad}
\address{Neural Computing Research Group\\
Aston University,Birmingham, UK}
\ead{jack.raymond@physics.org, saadd@aston.ac.uk}
\begin{abstract}

Code Division Multiple Access (CDMA) in which the spreading code
assignment to users contains a random element has recently become
a cornerstone of CDMA research. The random element in the
construction is particular attractive as it provides robustness
and flexibility in utilising multi-access channels, whilst not
making significant sacrifices in terms of transmission power.
Random codes are generated from some ensemble, here we consider
the possibility of combining two standard paradigms, sparsely
and densely spread codes, in a single composite code ensemble.  The
composite code analysis includes a replica symmetric calculation of
performance in the large system limit, and investigation of finite
systems through a composite belief propagation algorithm. A variety of codes
are examined with a focus on the high multi-access interference
regime. In both the large size limit and finite systems we demonstrate scenarios in which the composite code has typical performance exceeding sparse and dense codes at equivalent signal to noise ratio.
\end{abstract}

\maketitle
\section{Introduction}
Code Division Multiple Access (CDMA) is an efficient method of
bandwidth allocation, employed in many to one wireless
communication channels~\cite{Verdu:MD}. Schematically, each user
is allocated a code by which to modulate some sent symbol across
the bandwidth. The signal arriving at a receiver (base station) is
a superposition of the user signals and channel noise; with
carefully chosen codes, the source signals sent by each user may
be robustly inferred. The problem we wish to address is one of
multi-user detection, in which the bandwidth access patterns for
different users are random and not correlated in such a way as to
prevent, or reduce optimally, Multi-Access Interference (MAI).

The base station must extract information from the relevant parts
of the bandwidth in order to decode for a particular user. It is
convenient to consider two spreading paradigms. In the first, each
user transmits on the full bandwidth allocating a small amount of
power to each section. Alternatively, the user may have power
concentrated on one or several small sections of the bandwidth. In
the former case we say the code is  dense, and in the latter,
sparse. We will consider the case where the access patterns are
random and uncoordinated between
users~\cite{Tanaka:SMA,Yoshida:ASS,Raymond:SS,Guo:MDSS,Montanari:BPB}.
We also consider one simple case in which there is coordination
between users. Coordination between the users allows
opportunities to reduce MAI, thus producing an improved
performance.

In the CDMA with dense spreading patterns, bits of information may be transmitted at a
near optimal rate using pseudo-random dense spreading
codes~\cite{Verdu:MD}, which are amongst the best understood CDMA
methods. These codes are generated independently for each user, and
may be quickly decoded by a matched filter or modified message
passing methods under standard operating conditions. A more recent
interest has been in the sparse analogue of these codes, in which
performance is comparable but decoding is based upon Belief
Propagation (BP)~\cite{Montanari:BPB}. There exists enough
latitude in parameters and channel properties encountered in real
systems to anticipate that each method may be optimal in different
applications and operating conditions.

We argue that a principled simultaneous use of sparse and dense
codes might provide some useful diversity, and it is this
possibility which is investigated here. In particular, we consider
a linear combination of sparse and dense random codes to study
their generic properties.

We are also interested in the case of composite systems more
broadly than for the purpose of coding multi-access channels
~\cite{Raymond:OC,Hase:DA}. In statistical physics sparse random
graphs (representing sparse random codes) and fully connected
graphs (representing dense random codes) form the basis for many
statistical models. These models allow an analytical appreciation
of phenomena in physics and more general complex
systems~\cite{Nishimori:SP}. The CDMA model provides some
interesting indications of new phenomena when both sparse and
dense processes are at work.

\subsubsection{The model}
\begin{figure}[!t]
\centering
\includegraphics{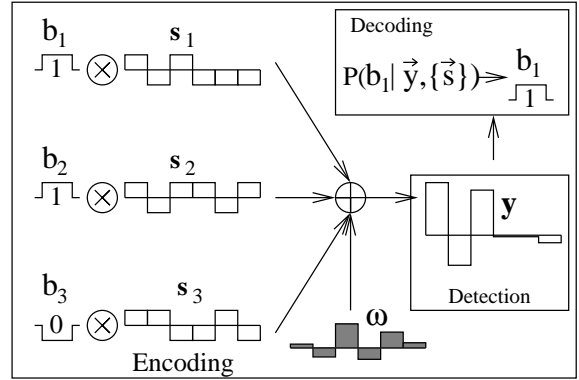}
\caption{\label{ss} During a bit interval each user $k$ sends a
bit $b_k$ modulated across some set of chips at constant amplitude,
either phase or anti-phase (a 2 symbol alphabet for each user).
The signal arriving at the base station is a superposition of user
signals and channel noise $\vomega$. The signal is detected and
decoded to recover estimates of the sent bits. In this diagram the
number of users $K=3$ and number of chips $N=5$, though in this
paper we consider cases where $N$ and $K$ are large. We omit the
fading factors for brevity from this description $F_{\mu k}=1$. }
\end{figure}

The model for wireless multi-user detection is idealised as a
linear vector multi-access channel subject to Additive White
Gaussian Noise (AWGN). The transmission between each user and the
base station is perfectly synchronised and power controlled. In
other words, the coding is received as intended at the detector,
there is no fading or scattering of the source signals.

The bandwidth is discretized as $N$ time-frequency blocks (chips),
so that a vector describes the spreading pattern across the
bandwidth. Each users (labelled by $k=1\ldots K$) is assigned a
modulating code ($\sk$) for transmission/detection of a random
bit, $b_k=\pm 1$ sent to/from a base station. The channel load is
$\chi=K/N$, which is finite. Consider the transmission case where
the base station has knowledge of all codes in use. A
superposition of the user transmissions, along with noise
($\vomega$) arrives at the base station as shown in figure
\ref{ss}. Assuming perfect synchronization of the user
transmissions with the chips the received signal on each chip
$\mu$ is
\begin{equation}
  y_\mu = \omega_\mu + \sum_k b_k F_{\mu k} s_{\mu k}\;. \label{channel}
\end{equation}
where $F_{k \mu}$ is the user specific fading, which may, in
practical applications, be different for each chip and user. In
order to allow good decoding the base station may coordinates the
amplitude of codes so that in expectation the received Signal to
Noise Ratio (SNR) is uniform for all users, which is a special
case of power control. For example, users at greater distances
(suffering greater fading) in a real cellular phone network will
be instructed to use a higher transmission power to mitigate this
effect. We assume such a determination of relative power levels
has been achieved, so that we may take $F_{k \mu}=1$ and
normalized codes $\vs_k^T \vs_k = 1$, the normalization can be
taken without loss of generality in the analysis since the noise
power spectral density can be rescaled accordingly. If the noise
distribution is independent and Gaussian distributed on each chip
then the SNR per bit is identical for all chips and defined as
\begin{equation}
 \SNRb = 1/(2 \sigma_0^2) \;,
\end{equation}
where $\sigma_0^2$ is the variance of the noise per chip. Note
that the total signal to noise ratio (the power spectral density)
contains an extra factor of $\chi$.

\subsubsection{Sparse, dense and composite spreading codes}

In standard dense CDMA a code is assigned to each user so that on
any chip the signal transmitted is modulated according to
$s^D_{\mu k}$, which is non-zero for most or all chips. The Binary
Phase Shift Keying (BPSK) random ensemble takes for each user a
normalized code sampled uniformly at randomly from $\{\pm
1/\sqrt{N}\}^N$.

In one definition of sparse CDMA, by contrast, there is no
transmission by user $k$ except on some finite subset
($\partial_k$) of $C_k$ chips ($\mu_1\ldots\mu_{C_k}$). Let $C$ be
be the mean connectivity users in the ensemble, which is small and
finite for the ensembles we study. The simplest case to consider
is one in which the set $\partial_k$ is sampled independently and
uniformly from the set of $N$ {\em choose} $C_k$ possible chip
combinations for each user. Throughout this paper we consider the
number of accesses for all users to be identical $C_k=C$. For the
subset of $C$ chips transmitted on by user $k$ ($\mu \in
\partial_k$), in the standard case, BPSK
modulation is used so that $s^S_{\mu k}$ is sampled uniformly from
$\{\pm\sqrt{1/C}\}^{C}$.

To assume the number of accesses per user is a random, sampled
from a Poissonian distribution, would also seem a natural choice,
but note that with this choice some users will fail to access the
bandwidth altogether. These disconnected users would strictly
limit performance, and constitute a fraction $\exp(-C)$ of all
users. Choosing the uniform distribution allows uniform access to
the bandwidth for all users in expectation, and we expect this to
be optimal in many senses~\cite{Raymond:SS}.

In both cases the modulation method is BPSK. This is not the only
viable modulation method, other simple modulation methods may
result in improved performance ~\cite{Raymond:Thesis}. One
convenient simplification in the case of BPSK is the symmetry of
the modulation, which means that for purposes of analysis the
transmitted bit sequence can be chosen as $\vb=\vones$ (all 1's)
without loss of generality.

\begin{figure}[!t]
\centering
\includegraphics[width=\linewidth]{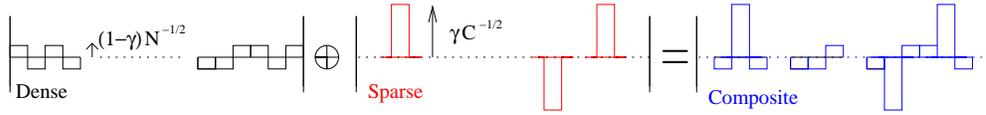}
\caption{(colour online) \label{sd} The upper figure shows a standard random BPSK
code for a dense system.  The middle figure shows the sparse
ensemble where all power is concentrated on a few ($C=3$) chips at
higher power on each chip. The composite system is a superposition
of these systems, the power in the sparse system is normalized to
$\gamma$ and in the dense code to $1-\gamma$. For the finite size
examples we avoid the collision between the two codes on $C/N$
links by setting the dense code to zero on the set $\partial_k$
and redistributing the power uniformly. In the large system limit
the overlap is in any case a negligible effect.}
\end{figure}

The codes we shall consider in this paper are composite codes
which may be constructed as a sum of a sparse and dense codes
sampled according to the definitions of this section. The
composite code is sketched in figure \ref{sd}, it involves both a
sparse and dense random codes with power normalized to $1$. The
power ratio between the sparse and dense parts is controlled by a
parameter $\gamma$, the spreading code may be written in this case
as
\begin{equation}
\vs^C_k = \sqrt{\gamma} \vs^S_k + \sqrt{1-\gamma} \vs^D_k\;.
\end{equation}
If the sparse and dense codes are normalized the new code will be
normalised up to a small ($O(1/N)$) factor which is not important
in the large system analyses. In the finite systems we correct for
this difference.

The difference between composite and dense codes is in the
hierarchical nature of the modulation sequences, all chips are
transmitted on, but with two scales of transmission (provided
$\gamma \gg 1/N$), so that in terms of information transmission
and decoding the subset of chips $\partial_k$ for user $k$ does
not become insignificant as $N$ becomes large, as would be the
case for any subset of the chips given a dense code for user $k$.

\subsubsection{The case for random codes}

Random codes, sampled according to some ensemble description,
offer flexibility in managing bandwidth access by allowing code
assignment by independently sampling for each user, and also have
robust self-averaging performance for large system sizes.
Furthermore the unstructured nature of codes makes them
useful in adversarial models of communication, or
in the presence of structured noise effects.

Random codes, sampled independently for each user, interfere in
the channel. Optimal encoding of sources would involve a
correlation of codes so as to minimise MAI. It has been shown that
standard dense and sparse random codes can achieve a $\BER$
comparable to optimal transmission methods in the AWGN vector
channel with only a modest increase in power. Optimal is here by
comparison with transmission in the absence of MAI, the single
user case, with comparable energy per bit transmitted. The small
increase in power required to equalise performance is often a
tolerable feature of wireless communication.

CDMA methods can be formulated so as to reduce or remove MAI; for
example orthogonal codes ($(\vs^*_k)^T \vs^*_{k'}=\delta_{k,k'}$)
can be chosen for sparse and dense systems, whenever $\chi\leq 1$,
achieving a single user channel performance. Gold
codes~\cite{Gold:OB} are the structured generalization of this
beyond $\chi=1$ in the case of the dense ensemble, where the
distance between codewords is maximized, but some MAI is present.
A sparse version of the orthogonal case is Time or Frequency
Division Multiple Access (TDMA/FDMA), whereby each chip is
accessed by at most a single user. For $\chi>1$ (but sparse,
$C=O(1)$) improvement is also possible beyond the random ensemble
-- for example by enforcing the chip connectivity to be
regular~\cite{Raymond:SS}.

In many cases only limited coordination of codes might be
possible, so that MAI is an essential and irremovable feature.
Random code ensembles are used in this paper; with random MAI present implicitly. The composite code
ensemble, like sparse and dense cases, has a structure which is
suitable to detailed mean-field type analysis in the
spread-spectrum limit, and as will be shown, can outperform sparse
and dense analogues in some reasonable parameterisations of the
linear vector channel.

The random coding models, with a little elaboration, may also
approximated different scenarios other than ones corresponding to
deliberately engineered codes. Consider for example a TDMA code,
which is a sparse orthogonal coding method, so that the
transmitted signal for any user is intended for a unique chip
(time slot). In a practical environment this code may not arrive
perfectly but might have a significant power component delayed by
random processes, which contributes to unintended chips. This may
occur in practice by way of multi-path
effects~\cite{Rappaport:900}. In terms of the optimal detection
performance, the properties may more closely resemble sparse CDMA,
rather than a MAI-free TDMA method. If the paths are more
strongly scattered across a significant fraction of the bandwidth
a random dense inference problem is implied. Finally, a scenario
with a few strong paths and many weak paths may apply, then the
detection problem might be best approached as something similar to
the composite code model.

\subsubsection{The case for a statistical treatment}
The large $N$ (spread-spectrum) scenario is an efficient
multi-user transmission regime~\cite{Ipatov:SS}, and one in which
we expect typical case performance of different codes drawn from
our ensemble to converge. In the large system limit
$N,K\rightarrow \infty$ the typical value for performance
statistics, that describe accurately almost all samples, are
the quantities of interest. Taking $C$ finite the
properties of dense, sparse and composite random codes are
distinguishable and may be calculated from a free energy density.
The spread-spectrum and many user limit is a standard benchmark,
and statistical mechanics methods are established tools in
analysis of such cases ~\cite{Guo:MDSP}. Often systems of
practical size reflect strongly the properties inferred from the
large system result; however, finite size effects may be critical
in some practical issues of decoding.

\section{Method}
\subsection{Calculation of Optimal Performance}

To determine optimal performance we analyze the mutual information
between the sent bits and received signal per bit $I(\vb,\vy;
\{\vs\})$. In the large bandwidth limit and for typical bit
sequences, channel noise and code samples performance will be
indistinguishable from the mean quantity
\begin{equation}
\se = \lim_{K\rightarrow\infty} \frac{1}{K}\Big\langle
-P(\vb,\vy)\log_2(P(\vb,\vy))\Big\rangle_{\vb,\vy,\{\vs\}}
\end{equation}
Where for our ensemble we will take $\vb=\vones$. This quantity is
expected to be self-averaging with respect to typical instances of
the code. Self-averaging assumes that cases differing
significantly from the mean value become statistically negligible
as $K=\chi N$ becomes large. The mean of the mutual information
with respect to these quantities is then a sufficient measure of
equilibrium properties.

This quantity may be most concisely determined through a
statistical mechanics methodology. The probability distribution
should be separated into a noise part, which is marginalised over,
and a bit estimate part, which is the part of interest
\begin{equation}
    P(\vb,\vy) \!=\! P(\vb)  \int \prod_\mu \left[\rmd \nu_\mu \delta\left(y_\mu\! - \!\sum_{k=1}^K s_{\mu k} b_k - \nu_\mu\right)\! P(\omega_\mu=\nu_\mu)\right]
    \;, \label{compositeCDMA.Pvbvy}
\end{equation}
For decoding purposes we assume an AWGN model, with signal to
noise ratio per bit estimated as $\beta \mQ$ ($\mQ$ abbreviating
SNR in equations), and marginalize with respect to this
distribution to find a quadratic form in the exponent. The assumed
form is within our model system correct, but in more complicated
systems the form may still be used as an estimate. The prior is on
bits is assumed to be uniform. The analysis then becomes
equivalent to evaluation of an Ising spin model with Hamiltonian
\begin{equation}
 {\cal H}(\vtau) = \mQ \sum_{\mu=1}^N \left(y_\mu -\sum_{k=1}^K s_{\mu k} \tau_k\right)^2\;,
\end{equation}
and inverse temperature $\beta$. The dynamical variables
$\{\tau_k\}$ approximate the random variables $b_k$, using the
evidence $y_\mu$ and the code $\Ms$ (quenched variables). The
mutual information is affine to the free energy density for this
model and takes an upper bound of $1$ bit.

The free energy may be evaluated through the replica
method~\cite{Nishimori:SP}. The self averaging free energy,
averaged over code samples and instances of the channel noise, may
be calculated by use of the replica identity
\begin{equation}
\fl \beta f\! = \!\lim_{K \rightarrow \infty} \<-\frac{1}{K}\log Z\>\!
=\! \lim_{K \rightarrow \infty} -\frac{1}{K}\lim_{n \rightarrow
0}\! \frac{\<Z^n\>\!-\!1}{n}\;; \qquad Z\!=\!
\sum_\vtau\exp -\beta H(\vtau)\;.
\end{equation}
To apply the method we take $n$ as an integer, with $Z^n$ becoming
a product of partition functions in different dynamical variables,
but shared quenched variables~\cite{Mezard:SGT}. For each of the
replicated partition functions a Gaussian identity may be applied
to reduce the square in the exponent to a linear form
\begin{equation}
\fl \exp\{ -\beta {\cal H}(\tau^\alpha_k)\} =  \prod_{\mu} \int  \rmd
\lambda_\alpha \exp[-\lambda_\alpha^2/2] \exp \sqrt{-\beta
\mQ}\lambda_\alpha \left(y_\mu-\sum_k s_{\mu
k}\tau^\alpha_k\right) \;.
\end{equation}
Representing $y_\mu$ as a product of the quenched variables
$\vb=\vones$,$\{\vs\}$ and $\vomega$ and then separating those
parts of order $1/\sqrt{N}$ in the exponent (due to the dense
code), we have
\begin{equation}
\fl y_\mu-\sum_k s_{\mu k}\tau^\alpha_k = \omega_\mu +
\sqrt{\gamma}\sum_{k \in
\partial \mu} s^S_{\mu k}(1-\tau^\alpha_k) + \sqrt{1-\gamma}
\sum_{k \setminus \partial \mu} s^D_{\mu k}(1-\tau^\alpha_k)\;.
\end{equation}
The sparse and dense code parts are now factorized in the exponent
and the quenched averages may be made independently according to
standard sparse and dense
methodologies~\cite{Raymond:SS,Tanaka:SMA,Raymond:Thesis};
introducing order parameters for the sparse and dense induced
correlations independently we have the free energy in a
variational form (\ref{eq:freeenrep}). The energetic ($\beta$
dependent) and entropic parts are separated for clarity, and the
entropy has been assumed extensive. Assuming a Replica Symmetric
(RS) form for the order parameters the free energy is determined by an extremisation
\begin{eqnarray}
\fl f \!&=&\! \Extr_{W,{\hat W},Q,{\hat Q},m,{\hat m}}\left[\frac{1}{\beta} f_s + f_e(\beta)\right]\;, \label{eq:freeenrep}\\
\fl f_s \!&=&\!  -\frac{Q{\hat Q}}{2} + m{\hat m} + \int \rmd W(x) \rmd {\hat W}(u) \log\left(\frac{1 + \tanh(u)\tanh(x)}{2}\right)\\
\fl\!&+&\! \frac{{\hat Q}}{2} +\int \prod_{c=1}^C \rmd{\hat W}(u_c) \< \log\frac{2 \cosh\left(\sum u_c + {\hat m} + \sqrt{{\hat Q}}\lambda\right)}{\prod 2\cosh(u_c)}\>_{\lambda} \;,\nonumber\\
\fl f_e &=& \!
-\frac{1}{\chi \beta} \!\log\left(\frac{\sigma_0^2}{A}\right)\!-\frac{1}{\chi\beta}\!
\int  \< \prod_{l=1}^L \left[\rmd W(x_l) \right] \right.\\
\fl &\times& \left. \< \log
\sum_{\tau_l} \frac{\exp\{x_l \tau_l\}}{2 \cosh(x_l)} \exp
\left\lbrace - \frac{1}{2 A}\left( \sqrt{A'}\lambda + \sum_{l=1}^L
\xi_l(1-\tau_l)\right)^2 \right\rbrace \>_{\{\xi_l\},\lambda}
\>_L\;,\nonumber
\end{eqnarray}
The order and conjugate (hatted) parameters are $W$ and ${\hat
W}$, normalized distributions over real fields, and $Q,{\hat
Q},m,{\hat m}$, which are scalars. The averages are with respect
to: $P(L)$, a Poissonian distribution (the excess chip
connectivity distribution of the sparse sub-code), $P(\{\xi_l\})$,
a uniform distribution on $\{\pm 1\}^{L}$ (the modulation
sequence), and $P(\lambda)$, a normally distributed variable.
Parameters
\begin{eqnarray}
 A \!&=&\! \sigma_0^2/\beta + \chi \left(1-\gamma\right) \left(1-Q\right)\nonumber\;,\\
 A'\!&=&\! \sigma_0^2 + \chi \left(1-\gamma\right) \left(1 - 2 m +Q\right)\nonumber\;.
\end{eqnarray}
describe the assumed and correct signal to interference [channel
noise and dense sub-code interference] ratios, respectively. If
$\beta$ is not taken as the Nishimori temperature, $\beta=1$, then
$A \neq A'$.

The sparse code requires a full distribution to describe the
contribution to the free energy, whereas the dense part requires
only a few scalar parameters. Taking the limiting cases of large
or small $\gamma$ one set of order parameters becomes negligible
and the usual sparse/dense expressions are recovered
~\cite{Raymond:SS,Tanaka:SMA}. The complexity of evaluating the
free energy is limited primarily by the sparse part.

At the Nishimori temperature, corresponding to the correct
detection model ($\beta=1$), the free energy is correctly
described by the RS assumption~\cite{Nishimori:CO}. With this
assumption we can make use of a symmetry to reduce the number of
order parameters in the dense part ($Q=m$,${\hat Q}={\hat m}$).
Extremization of the free energy leads to the following sets of
saddle-point equations in the order parameters.
\begin{equation}
\begin{array}{lcl}
\fl W(h) \!&=&\! \int \prod_{c=1}^{C-1} \rmd u_c {\hat W}(u_c)\delta\left(h -\sum_{c=1}^{C-1} u_c + {\hat Q} + \sqrt{{\hat Q}}\lambda\right) \\
\fl{\hat W}(u) &=&  \int \< \prod_{l=1}^L \rmd h_l W(h_l)\delta\left(u - \frac{1}{2}\sum_{\tau_0} \tau_0 \log Z'(\tau_0)\right)\>_{L,\{\xi_l\},\lambda} \\
\fl Q \!&=&\! \<\int  \prod_{c=1}^{C} \rmd u_c {\hat W}(u_c) \tanh^2\left(\sum_{k=1}^C u_k + {\hat Q}+\sqrt{\hat Q}\lambda\right)\>_{\lambda}\\
\fl {\hat Q} \!&=&\! \chi \left(1-\gamma\right)(1-Q)
\end{array}\label{saddlepointeqn}
\end{equation}
where the distributions by which averages are taken are unchanged.
The quantity $Z'$ is a local mean field partition sum, only the
dependence on one summation variable is written explicitly,
\begin{equation}
Z'(\tau_0)\!=\!
\prod_{l=1}^L\left[\sum_{\tau_l}\right]\!\exp\!\left(\!\sum_{l=1}^L
\!h_l \!\tau_l \!-\!  \frac{1}{2}\!\left(\!\lambda \!+\!
\sum_{l=0}^{L}\! \frac{\xi_l}{\sqrt{A}}\!
(1\!-\!\tau_l)\!\right)^2\!\right)\;.
\end{equation}
Equations are easily modified to include the general $\beta$ case.

From the free energy at the saddle-point, by application of small
conjugate field against the terms $\sum_\orderedL{i}
\tau_{i_1}\ldots \tau_{i_L}$, it is possible to identify $P(H)$
with the distribution of log-posterior ratios of source
reconstruction in typical instances of the quenched model. Let
\begin{equation}
H_k = \frac{1}{2}\log \left(P(b_k=1| \Ms,\vy )/P(b_k=-1 | \Ms,
\vy)\right)
\end{equation}
then the quantity
\begin{equation}
\fl P(H) = \lim_{K\rightarrow\infty} \frac{1}{K}\sum_{k=1}^K \delta (H
- H_k) = \int \rmd u \rmd h {\hat W}(u) W(h) \delta(H - (u+h))
\label{eq:compositeCDMA.PH}\;.
\end{equation}
once an analytic continuation is taken in the sum. From this
observation the bit error rate is determined by the integral
\begin{equation}
\BER = \int_{-\infty}^0 P(H) \rmd H \;.
\end{equation}
The mutual information, with regards to typical case of
(\ref{compositeCDMA.Pvbvy}), is attained by an affine
transformation of the free energy. The entropy is also calculated
from the same treatment, at the Nishimori temperature the energy
is $1/(2\chi)$.

\subsection{Decoding: multistage detection and BP}
The idealized achievable performance is calculated in the limit of
large $N$ under the RS assumption. In practice one must deal with
finite systems, and the finite size effects tend to degrade
performance relative to the ideal. However, for reasonable size
systems ($N \gtrsim 100$) and $\gamma\gg1/N$ the properties of
composite codes in decoding, based on suitably constructed
heuristics, become distinguishable from the performance through
sparse or dense decoding methods, and approach in many cases the
solutions predicted by the replica method.

We consider two algorithms, the standard multistage detection
(MSD)~\cite{Verdu:MD}, based on iteration of a vector
approximation to the sent bits
\begin{eqnarray}
\vb_k^{t+1} \!&=&\!  \sign\left[(\vs_k)^T \vy -  \sum_{k'\setminus k} Y_{k k'} b_{k'}^{t}\right] \label{MSD}\\
 Y_{k k'} \!&=&\! \vs_k . \vs_{k'} \nonumber
\end{eqnarray}
and a modified form of BP. MSD is a heuristic method
~\cite{Verdu:MD}, which works well in dense codes and simple noise
models, provided MAI is not too large. BP is based on passing of
conditional probabilities (real valued messages) between nodes in
a graphical representation of the problem~\cite{Kschischang:FG} (a
schematic description of the method appears in figure
\ref{BPgraph}).

The most time consuming step in BP is a trace over the states
attached to a particular factor node in order to determine the
evidential message, a naive approach in the dense case requires
$O(2^N)$ floating point operations. However, due to the central
limit theorem the dependence on the weakly interacting bits, not
connected strongly through the sparse code, is equivalent to a
Gaussian random variable and the marginalization is replaced by an
exact Gaussian integral. This reduces algorithm complexity
asymptotically to $O(N^2)$.

The approximation leads to a more concise form for the {\em
evidential} messages (passed from factor nodes to variable nodes):
\begin{eqnarray}
\fl u^{t}_{\mu \rightarrow k} \!&=&\! \sum \tau_k \frac{1}{2 \beta}\log\left( Z_{\mu \rightarrow k}(\tau_k)\right)\;,\label{cavbiasmessS}\\
\fl Z_{\mu \rightarrow k}(\tau_k) \!&\doteq&\! \prod_{l \in \partial_\mu \setminus k} \left[\sum_{\tau_l}\exp\left\lbrace \beta h^{t}_{l\rightarrow \mu} \tau_l \right\rbrace\right]  \\
\fl &\times&\exp\left\lbrace - \frac{1}{2 A^{t}_{\mu k}}\left(y_\mu - \sum_{l \in \partial_\mu} s_{\mu l}\tau_l - \sum_{l\setminus \partial_\mu} s_{\mu l} \tanh(\beta H^{t}_l)\right)^2\right\rbrace\;, \label{Zmuk}\\
\fl A_{\mu k}^{t} \!&=& \!\frac{\sigma_0^2}{\beta} \!+ \!\!\!\sum_{l\setminus \{k ,\partial_\mu\}} \!\!\! s_{\mu l}^2 \tanh^2 (\beta h^{t}_{l \rightarrow \mu}) \!\doteq \!  \frac{\sigma_0^2}{\beta} \!+\! \chi (1-\gamma)\left(1 \!-\!\frac{1}{K}\sum_{l=1}^K \tanh^2 (\beta H^{t}_{l})\right)\;, \label{Amuk}
\end{eqnarray}
where a further simplification is possible for messages passed along dense links, using an expansion to leading order in $s_{\mu k}= O(1/\sqrt{N})$,
\begin{equation}
 \fl u^{t}_{\mu \rightarrow k} \!\doteq\! \frac{1}{\beta A_{\mu k}^{t}}
s_{\mu k}\left(y_\mu - \sum_{i\setminus \{k ,\partial_\mu\}}
s_{\mu i} \tanh (\beta H^{t}_i) - \sum_{l \in \partial_\mu} s_{\mu k}
\tanh (\beta h^{t}_{l \rightarrow \mu})\right)\qquad  \label{cavbiasmessD} \;,
\end{equation}
as constructed in ~\cite{Mallard:BPDG}. In these expressions the
notation $\doteq$ indicates those equations where some $O(1/N)$
corrections have been eliminated, the most critical being the
replacement of the full marginalization over densely connected
variables in (\ref{Zmuk}) by a Gaussian integral that is taken
analytically. At termination time the set of bits is determined by
$\vb = \sign(\vH^{(T)})$. Evidential messages may be combined in a
standard way to give marginal log-posterior estimates for the
source bits
\begin{equation}
H^{t+1}_k \!=\!\frac{1}{2 \beta}\log \frac{P(b_k=1|
\vy)}{P(b_k=-1| \vy)} = \sum_{\mu=1}^N u^{t}_{\mu \rightarrow k}
\;,
\end{equation}
 and {\em variable} messages (passed from variable nodes to factor nodes)
 \begin{equation}
 h^{t+1}_{k \rightarrow \mu} \!=\! H^{t+1}_k - u^{t}_{\mu
\rightarrow k} \;.\label{cavfieldmess} \;.
\end{equation}
The algorithm remains $O(N^2)$, and is less convenient than
(\ref{MSD}); however, the expression may be manipulated without
introducing any additional errors at leading order in $N$ to an
algorithm without dense messages~\cite{Raymond:Thesis}, the
manipulation is an application of that proposed in
\cite{Kabashima:SMA}. The asymptotic complexity of the algorithm
is comparable to MSD and other good algorithms, although this
disguises a large constant factor and some structures which might
be difficult to implement efficiently in hardware.

Composite BP is applied as a heuristic algorithm based on an
unbiased initialisation of the messages, in the hope that the
various simplifications on the algorithm do not produce strong
finite size effects. BP exactly describes the marginal probability
distributions only if the messages (\ref{cavbiasmessS}),
(\ref{cavbiasmessD}), (\ref{cavfieldmess}) converge to a unique
fixed point, since in this case $\vH$ describe the log-posterior ratios.
There are two scenarios to be concerned about, either the BP
messages fail to converge, or they converge to an incorrect fixed
point - both scenarios occur in different decoding regimes for
CDMA. The requirements for standard BP to successful decode are
closely related to the assumption of RS, hence the similarity of
the minimization process for $W,{\hat W}$ (\ref{saddlepointeqn})
and the BP equations.

In studying systems of finite size we consider two message update
schemes for both MSD and BP. The first is a parallel update scheme
where all variables are updated such that the values of the
current generation of messages ($t+1$) are conditionally
independent given the previous generation of messages $t$. The
second schemes is a random stochastic update method, the updates
are applied to all messages in the population, but in a random
order. As soon as a variable is updated it is made available to
subsequent updates, the messages in a single generation ($t+1$)
are then not conditionally independent given the previous
generation ($t$). The sequential update method is slower to
implement, but helps to suppress oscillations observed in some
parallel update schemes, that lead to unphysical solutions.

A measure of convergence for the BP  is the mean square change in
variable states
\begin{equation}
 \lambda^{t}=\frac{1}{K}\sum_{i=1}^K
 \left(H^{t}_i-H^{(t-1)}_i\right)^2\;,\label{eq:lambda}
\end{equation}
whereas the convergence measure for MSD is similar but with the
log-likelihood ratios replaced by the bit estimates. An
exponential decay in this quantity, or an evaluation to zero,
would be characteristic of a convergent, or converged, iterative
method.

\subsection{Properties of decoders in finite systems}
MSD is an iterative method which works very well in systems with
small load $\chi$ and mixing parameter $\gamma$. In the first
iteration the achieved result is equivalent to a matched filter.
In subsequent iterations the estimates are updated, but because
the information is rather crudely used the consequence can be
instability of the iterative procedure. For instance, the examples
shown later reveal very poor performance of MSD when MAI is large.
Since MSD is based on filtering it is not so successful for
composite ensembles as for dense ones, and its reliability in
dense codes improves as system size increases.

\begin{figure}
\includegraphics[width=\linewidth]{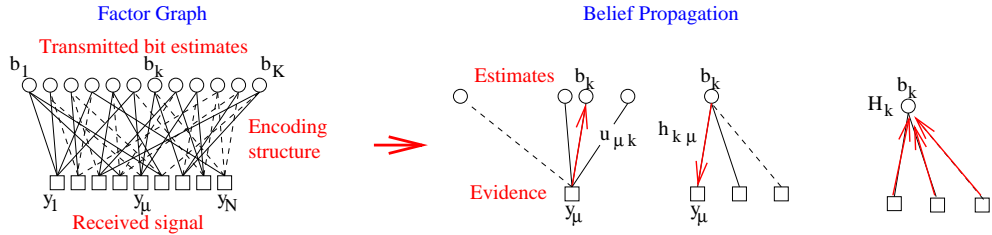}
\caption{(colour online) \label{BPgraph} Left figure: The factor graph
representation of the problem is useful, especially in visualizing
the topology of sparse problems. Right figure: BP based on updates
of link variables is often successful in determining a probability
distribution consistent with the evidence, for both sparse and
dense systems. The messages passed are represented by arrows and
are conditional probabilities in BP, other message passing schemes
are also possible, a special case being multi-stage detection.}
\end{figure}

The critical scenario in which BP is guaranteed to produce the
correct marginal posteriors is that the graphical model is tree
like (see figure \ref{BPgraph}). However, BP often produces a
reasonable performance in loopy models including
sparse~\cite{Montanari:BPB} and dense~\cite{Kabashima:SMA} CDMA.
The failing regime in BP corresponds to large $\chi$ in the sparse
and dense codes, but at intermediate $\gamma$ the instabilities
appear to be worse, or new instabilities are apparent.

In many of the cases studied we found that BP did converge in the
marginal log-likelihood ratios, this was the case for systems at
small $\chi$, and/or high SNR. In other cases the fields did
not converge, and instead a steady state was reached at the
macroscopic level -- remembering the BP equations describe a
dynamic algorithm, which does not obeying detailed balance, this
might be expected. The steady state is one in which the
distribution of messages converges up to finite size effects, but
the individual messages do not converge. Steady states were
characteristic of systems initiated with random or unbiased
messages at high $\chi$. The estimates determined from the
distribution of messages in the steady state typically corresponds
to a high $\BER$ estimates.

In regimes where message passing is unstable the detectors may
still be used to provide an estimated, subject to some termination
criteria. Variations on MSD and BP involving heuristic tricks may
avoid some of these effects, but some of the standard methods may
be unsuitable to the composite model. Experimenation with the
update scheme demonstrated improved results in MSD for example.

The dynamics of numerically solving the saddle-point equations
(\ref{saddlepointeqn}) are very closely related to the iteration
of BP. The dynamics of the iterations (figure
\ref{metatimeseries}) appear smooth and systematic even at large
$\chi$ based on a numerical solution involving $10000$ points.
However, in addition to the use of a large system size, control
was exercised over finite size effects through selective sampling
of quenched variables in the update step (\ref{cavbiasmessS}),
which is not possible in BP. Therefore any realisation of the
problem in BP, even at an equivalent system size, is not expected
to produce such uniform effects. Many qualitative features such as
the speed of convergence appear to be
reproduced in some of our finite realisations.

Naturally we expect there to be finite size effects, the trends
presented appeared consistent across a range of system size from
$O(100)$ to $O(1000)$ chips. No structured attempt is made to
calculate these effects, or to distinguish the contributions due
to the different $O(1/N)$ approximations in the algorithm and
other instabilities implicit to BP. Working with a sufficiently
large graph is problematic since the algorithm is asymptotically
$O(N^2)$ rather than linear, and it is necessary to store and
manipulate a $K \times N$ modulation pattern matrix.

\subsection{Replica symmetry and the phase space}

At the Nishimori temperature the RS solution is guaranteed to
describe correctly the thermodynamically dominant states. Across a
range of parameters we find two saddle-point solutions: one
corresponding to a locally stable {\em bad solution} (bad decoding
performance) and one to a locally stable {\em good solution} (good
decoding performance). The metastable solution (solution producing
the higher free energy) is irrelevant in determining the
equilibrium properties since it does not contribute to the free
energy at order $N$. However, in terms of dynamics or local
sampling the bad solution can be dominant even where it does not
describe the equilibrium phase. We find these metastable regimes
at intermediate SNR and high MAI (above $\chi=1.5$
~\cite{Tanaka:SMA}).

The RS solution obtained at equilibrium appears sufficient to
describe the  good equilibrium and metastable solutions. These are
the local solutions to the free energy that correspond to states
clustered about the encoded bit sequence $\vb$. In this case the
phase is connected in state space, and so we expect the dynamics
of the system to be relatively simple, so that the phase space can
be explored by local sampling methods such monte-carlo. BP will be
locally stable in the vicinity of this solution, in the absence of
competing local minima we can expect convergence towards this in
specific realisations. This solution to the free energy exists
when SNR is sufficiently large.

By contrast we expect there to also be a bad (liquid/paramagnetic)
equilibrium solution when SNR is small. The field term in the
Hamiltonian means the magnetisation is never zero, but we expect
there to be a suboptimal ferromagnetic solution which is also
connected in state space, and that has similar properties in terms
of BP and sampling. The good and bad equilibrium solutions are
guaranteed to have simple phase space structures described by RS,
this conclusion and some thermodynamic properties can be
calculated without the replica trick or cavity
method~\cite{Nishimori:CO}.

Finally we can anticipate that at high MAI and intermediate noise
there may be a bad metastable solution. The bad metastable
solution emerges continuously from the bad equilibrium solution
with increasing SNR and so will be characterised by a connected
phase space for some parameters. However, as the noise decreases
we might expect this solution to become fragmented and the RS
metastable solution to become unstable, but this instability is
not tested. An indication of the failure of RS is the negative
entropy in some metastable solutions, which is not viable. It is not uncommon for systems with simple
connected phase spaces to exhibit negative entropy when the RS
ansatz is applied under the assumption of extensive
entropy~\cite{gross}, as we employ in the calculation. A result
without negative entropy can be formulated by a minor variation on
the RS approach called frozen replica symmetry breaking (RSB), which effectively rescales the
temperature. However, it is not certain that this solution will be
correct without a local stability analysis towards other forms of
RSB, in many other systems negative entropy is one indicator of a
failure of the connected phase space assumption~\cite{Mezard:SGT}.

In the bad metastable state, and also in the bad equilibrium phase
away from the Nishimori temperature ($\beta>1$), the connected
description is possibly incorrect even where the entropy is
positive; an RSB formalism may be
applicable. The good solution is likely to be well described by RS
at all temperatures, since it is a state clustered around the
encoded bit sequence.
For $\beta>1$ the RS approximation produces a variational
approximation to the thermodynamic behavior. The RS approximation
can also describe exactly the metastable states in some regimes,
but this is not the case in general.

The hypothesis of a connected state described by the RS treatment
has consequences for dynamics, as do the various hypotheses on the
nature of RSB, should it occur either in a search for the ground
state~\cite{Krzakala:GS}, or at some intermediate temperature. On
specific realizations of graphs we might expect BP to converge for
the states correctly described by RS. However, in finite graphs we
might expect strong finite size effects to dominant behaviour, so
that in the absence of a scaling analysis conclusions cannot be
drawn directly from simulation results. In cases where BP is
unstable due to RSB or finite size effects we might nevertheless
expect to uncover a steady state of the dynamics that is strongly
correlated with nearly optimal solutions and is useful for
extracting an estimate of sent bits.

\section{Results}
\subsection{Parameters considered}

The model constructed is already quite simple, avoiding some of
the practicalities of real channels and making no attempt to
optimize for finite size effects in the composite ensembles.
However, even with these simplifications the channel produces
interesting behavior. In order to demonstrate the equilibrium
properties of composite codes in such a way as to produce strong
contrast between the ensembles we work with systems with $C=3$ and
$\chi$ between $3/5$ and $2$. Beyond this range of $\chi$ results
appear by experimentation to be very much a continuation of the
trends highlighted.

Analysis of the sparse code ($\gamma=1$) is for this range of
parameters a loopy inference problem, but is sufficiently far from
the percolation transition to be well behaved. At the same time
$C=3$ is sufficiently small to allow quick decoding and produce a
contrast with the dense code. It has been noticed since the early
days of studying sparse codes that the mean connectivity of the
sparse code ensemble $C$ does not have to be very large at all for
results to become indistinguishable from the dense
code~\cite{Yoshida:ASS,Montanari:ABP}.

A lower bound to the achievable bit error rate in all ensembles is
given by the single user Gaussian channel (SUG) result over a bit
interval
\begin{equation}
 \hbox{SUG}= \int_{-\infty}^{0} \rmd \nu \frac{\sqrt{\mQ}}{\sqrt{\pi}} \exp\left\lbrace -\mQ (\nu-1)^2 \right\rbrace\;,
\end{equation}
which is the complementary error function of SNR. In the
absence of MAI this lower bound can be achieved if spreading
patterns are coordinated so as to be orthogonal. On the vector
channel this orthogonality is possible only if $K \leq N$,
unavoidable MAI at higher loads will strictly degrade performance.

The saddlepoint equations (\ref{saddlepointeqn}) are solved by
population dynamics~\cite{Mezard:BLSG}, an iterative method for
solving the saddle-point equations using a histogram approximation
of the distribution $W$ ($10000$ points are sufficient to attain
our results). Evolving the order parameters from initial
conditions that correspond to low and high $\BER$ finds either the
unique solution or a pair of locally stable solutions.

\subsection{Equilibrium behavior of unique saddle-point solutions}
\begin{figure}[!htbp]
\includegraphics[width=0.7\linewidth]{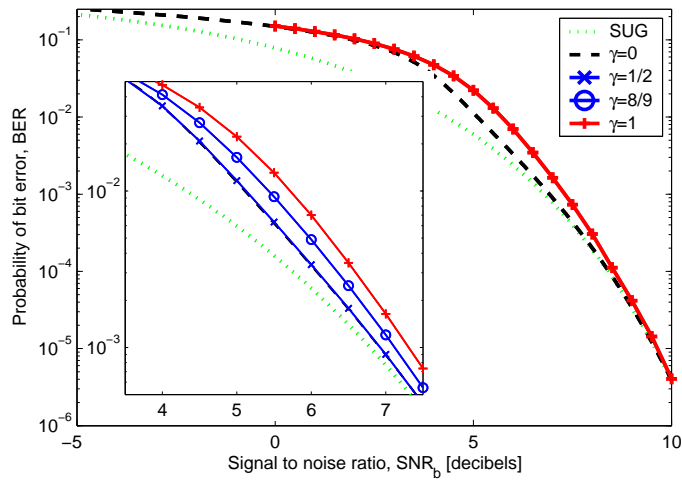}
\caption{(colour online) \label{normalregime} The figure demonstrates the $\BER$
determined from the order parameters at the equilibrium solution
of the free energy for various SNR and $\chi=1$. The curves
represent different ensembles ($\gamma$), with the single user
Gaussian (SUG) channel lower bound also displayed (dotted line)
for comparison. Error bars are significantly smaller than symbol
size for $\BER$ above $10^{-4}$, and are excluded for clarity. The
lower bound is approached for the CDMA codes at large and small
SNR, the dense code is best amongst the random codes. The code
with an even power distribution between the sparse and dense parts
($\gamma=1/2$) is not easily distinguishable in thermodynamic
performance from the dense code, even where the spread of codes is
greatest (inset).}
\end{figure}
Generally with $\chi \lesssim 1.5$ there is a unique solution of
the saddle-point equations with a smooth transition between bad
and good solutions as SNR is increased. The population dynamics
equations require very few iterations and results can be achieved
with relatively fewer points in the histogram. The normal working
range of CDMA is often by design one with a relatively small load
($\chi<1$) and so falls into this class of behaviour.

The trends in $\BER$ for the large system limit for $\chi=1$ are
demonstrated in figure \ref{normalregime}. The dense code ensemble
achieves a smaller bit error rate than the sparse code ensemble,
and the composite code ensembles interpolate between these. With
$\gamma=0.5$ the curve is indistinguishable at this magnification
from the dense curve, with balanced power in the two codes
performance resembles the dense code. At intermediate SNR there
is a large gap in BER, which narrows in the limits of high and small
SNR. Trends in the free energy follow a similar monotonic
pattern -- the dense code has the highest mutual information
everywhere.

\subsection{Metastable solutions of the saddle-point equations}
\begin{figure}[!htbp]
\includegraphics[width=1\linewidth]{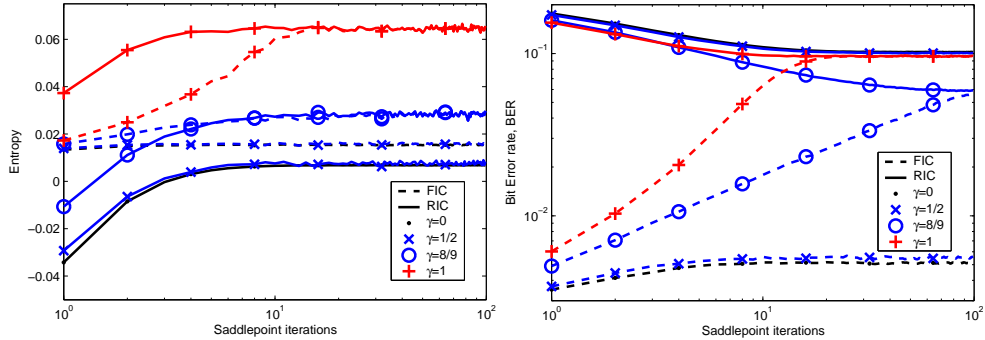}
\caption{(colour online) \label{metatimeseries} The dynamics of the order
parameters determined by iteration of the saddlepoint equations is
shown for $\chi=5/3$ and $\SNRb=6dB$, with a large population of
$10^6$ points to represent the distribution $W$
(\ref{saddlepointeqn}). Evolving the saddlepoints from either
Ferromagnetic or Random Initial Conditions (FIC/RIC) discovers
either the unique solution ($\gamma=1$ or $8/9$), or two locally
stable solutions ($\gamma=0$ or $1/2$). Left: The maximum free
energy is determined by the system of maximum entropy at the
Nishimori temperature. In cases of small $\gamma$ there are two
candidate solutions. The fluctuations are visible in some curves
and are due to the sampling method, these fluctuations are not
sufficient to escape the local solutions in the cases of
metastability. Right: $\BER$ is widely spread, for small $\gamma$
the thermodynamic solution is the good solution in this example.
At larger $\gamma$ there is a unique solution which is
intermediate between the metastable and thermodynamic solutions at
small $\gamma$.}
\end{figure}
\begin{figure}[!htbp]
\includegraphics[width=1\linewidth]{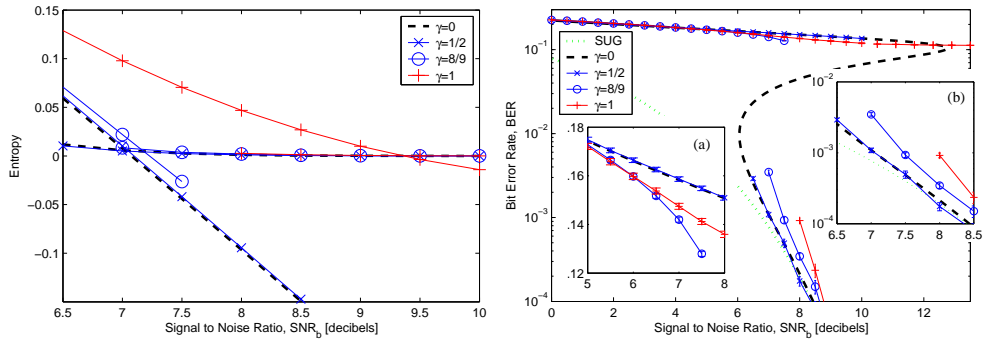}
\caption{(colour online) \label{metaregime} The figure covers the same range of
parameters as figure \ref{normalregime}, but with a load $\chi=2$.
Two locally stable solutions are found by minimisation of the RS
saddlepoint equations in a range of SNR for all $\gamma$. Left
figure: The entropy indicates a second order transition between
the good and bad solutions for each ensemble. At SNR greater
than the thermodynamic transition point metastable solutions
evolve towards a freezing point ($s=0$) and a regime of negative
entropy. The thermodynamic transition point is at significantly
greater SNR in the sparse ensemble than the composite
ensembles. The range of SNR for which metastability exists is
minimised in composite systems with $\gamma\approx 8/9$. Error
bars are everywhere much smaller than symbol size. Right figure:
The thermodynamic transition indicates a large gap between the
saddlepoint solutions at the thermodynamic transition and spinodal
points. The properties of the good and bad solutions change
smoothly about the thermodynamic transition and freezing point.
Right inset (a): The bad solution has high $\BER$ even at large
SNR and becomes locally unstable at lower SNR for composite
systems with $\gamma\approx 8/9$. Right inset (b) Good solutions
with larger $\gamma$ have lower $\BER$ and are stable at smaller
SNR. }
\end{figure}
The regime of high $\chi$ is of greater theoretical interest in
multi-user detection since this is where MAI causes results to
differ substantially from single user models. As $\chi$ is
increased beyond $1.5$ a spinodal point may be reached beyond
which there are multiple locally stable solutions to the
saddle-point equation.

In regimes with a competition between locally stable attractors,
or with one marginally stable attractor convergence of the
saddlepoint equations is slowed down; one such scenario is shown
in figure \ref{metatimeseries}. At $\chi=5/3$ there is a unique
solution for some of the sparse and composite ensembles, but not
for the dense ensemble. In this example the composite code
solution is superior to the sparse solution, and the dense
metastable (bad) solution. The best solution is the dense
thermodynamic (good) solution. As shown in figures
\ref{metaregime} and \ref{metatimeseries}, the entropy is positive
for all the thermodynamic solutions. However, at larger $\chi$ and
higher SNR the metastable solutions can have negative entropy,
indicating an inadequacy in the RS description.

The saddle-point solutions for our ensembles with load $\chi=2$ at
a range of SNR is shown in figure \ref{metaregime}. For this
load metastability is present at all $\gamma$ values. Where the
solution is not unique the correct and metastable solutions can be
distinguished from the free energy (equivalently entropy at the
Nishimori temperature). At the Nishimori temperature there is a
second order transition, the energy is equal to $1/2$ in both
solutions, which is realized as a discontinuous transition in the
BER. In the metastable regimes we find the entropy evolves towards
a negative value as SNR increases, the correct metastable state
in the negative entropy regime is described by the state space at
the freezing point where entropy first becomes negative.

Up to 7dB the bad solution is the thermodynamic solution in all
ensembles. Close to the transition the best performing codes are
composite ones with $\gamma\sim 8/9$, but at lower SNR the
regular code ensemble appears best. The composite systems
displayed all have thermodynamic transitions near 7dB, the entropy
and free energy of the sparse bad solution is much larger, so that
thermodynamic transition does not occur until about 9.5dB. This
entropy gap is a clear feature of the local freedom explicit in a
sparse connectivity model and absent in the composite one. In the
case of a regular sparse part, with a more homogeneous interaction
structure, the gap in entropy and thermodynamic transition point
are significantly reduced~\cite{Raymond:SS}. Amongst the good
solutions, in contrast to the bad solutions, both the ensemble
entropy and $\BER$ appear to be ordered by $\gamma$ for all
SNR.

The metastable solutions appear to be qualitatively similar in the
composite ensemble to the sparse and dense ensembles
~\cite{Kabashima:SMA,Raymond:SS}. What is interesting in the
metastable regime is that the positioning of the composite state
performance is not a simple interpolation between the sparse and
dense ensemble results. In the example shown the metastable
solutions for composite codes are at lower $\BER$ than either the
sparse or dense metastable solutions. Furthermore, for
$\gamma=8/9$ there is a unique solution beyond 8dB inspite of the
persistence of metastable solutions in the sparse and dense
ensembles at significantly larger SNR.

We did not test
the microscopic stability of the metastable solutions for the
composite system, but this should be possible, in part, by a local
stability analysis of the RS description.  It is expected that at,
and above, the Nishimori temperature ($\beta<1$) the RS
description will be locally stable even for the metastable states,
as was found for the dense and sparse
codes~\cite{Kabashima:SMA}~\cite{Raymond:SS}.

The composite codes exhibit a thermodynamic behaviour most
strongly contrasting with both sparse and dense codes when $\gamma
\lesssim 1$, and close to the thermodynamic transition of the
dense code. The effect of distributing power mostly in the sparse
code appears to destabilise the bad solution in some marginal
cases.  Results obtained by the replica method solved by
population dynamics appear robust; only close to the dynamical
transition points, where solutions are marginally stable, are not
well represented.

To understand the origins of this instability requires a more
detailed investigation of the stability of the RS metastable
solutions, and possibly an RSB type treatment. We suspect that the dense code
behaves comparably to the external field at equilibrium, and this
might be one way in which one could approach the problem.

\subsection{Decoder Performance}
\begin{figure}[!htbp]
\centering{
 \includegraphics[width=0.9\linewidth]{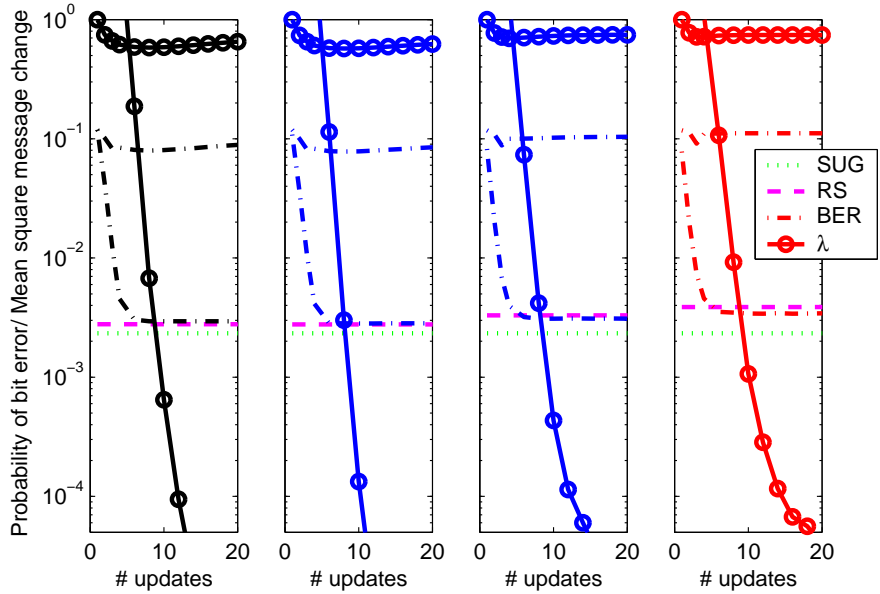}
\caption{(colour online) \label{normal_decoding} Mean $\BER$ (dashed line) and
$\lambda$ (solid line) are shown for different ensembles,
$\gamma=\{0,1/2,8/9,1\}$ from left to right, as a function of the
number of variable estimate updates for BP and MSD implemented
with parallel updates. $\SNRb=6dB$ and $\chi=3/5$ ($N=1000$,
$K=600$): for each point $300$ independent sparse and dense
connectivity profiles were sampled and combined in proportion to
$\gamma$, with channel noise randomly sampled from a Gaussian
distribution. The convergence measure $\lambda$ (\ref{eq:lambda})
indicates exponential convergence in BP and non-convergence of MSD
for all ensembles. The RS result is approached after $10$ updates
by the simulation average, but with some systematic error due to
finite size effects. The MSD result does not improve beyond about
five updates, but mean statistics reflect primarily the performance of the worst samples.}
}
\end{figure}
\begin{figure}[!htbp]
 \includegraphics[width=0.6\linewidth]{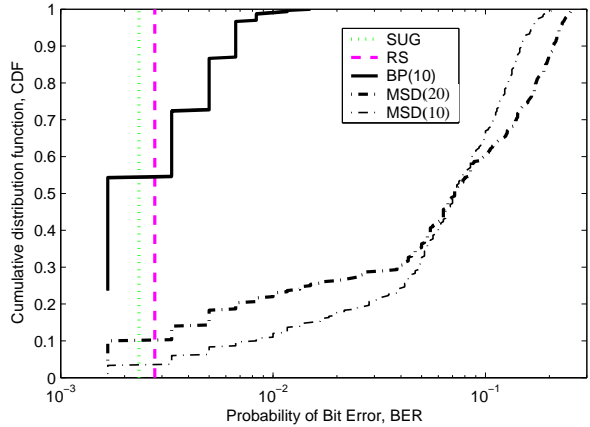}
\caption{(colour online) \label{normal_decoding2} The cumulative distribution
function for the decoding at $\gamma=0$ of the $300$ samples
taken, as in figure \ref{normal_decoding}, is typical in structure
of all composite systems. The $\BER$ found by BP has converged for
all samples taken within 10 updates. The $\BER$ found by MSD
continues to evolve between 10 and 20 updates, with increasing
$\BER$ for some subset of the samples. The median of the samples
is close to the BER, but the cumulative distribution function is
not yet approaching a tight Gaussian and finite size effects are
thus important. Some percentage of samples obtain a zero bit error
rate which accounts for a small part not represented on the
logarithmic scale.}
\end{figure}
\begin{figure}[!htbp]
 \includegraphics[width=0.8\linewidth]{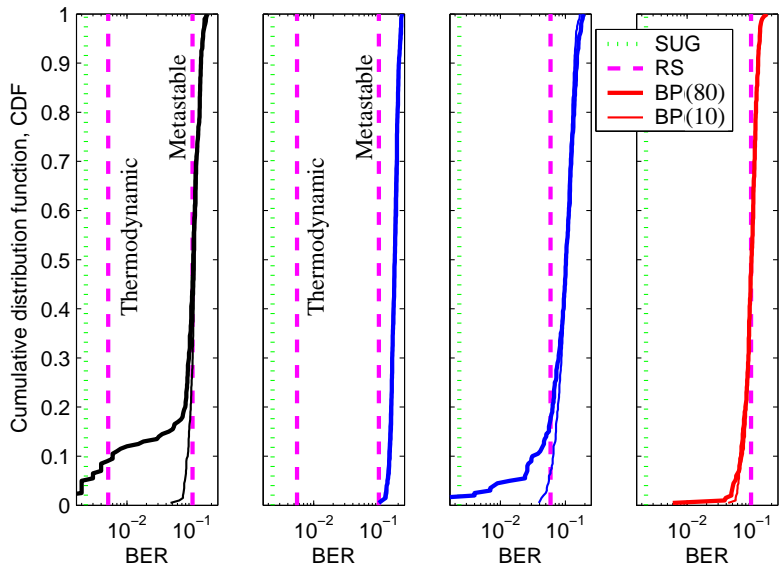}
\caption{(colour online) \label{decodingbadly} For $\SNRb=6$ decibels and $\chi=5/3$
($N=600$, $K=1000$) we demonstrate the decoding performance of
some algorithms by presenting cumulative distribution functions in
the $\BER$ based on $200$ runs. The histograms from left to right
represent mixing parameter values $\gamma=\{0,1/2,8/9,1\}$. The
sparse code samples (far right) converge in most cases after 10
iterations, and the median performance is close to the unique RS
solution. The dense ensemble (far left) is after 10 iterations
close to the median performance for the metastable RS solution
(right RS soluton). A subset of samples evolve further towards or
beyond the thermodynamic RS solution (left RS solution). For
$\gamma=1/2$, and $\gamma=8/9$, performance for most samples does
not reach the asymptotic RS prediction for the metastable solution
$\BER$.}
\end{figure}

In systems with $\chi \lesssim 1$ the equilibrium results are
achievable by iteration of BP equations, this was established
previously for the dense case in \cite{Kabashima:SMA}. Such an
example is shown in Figure \ref{normal_decoding} with $\chi=3/5$.
The performance of MSD is poor, although initially the achieved
bit error rate is improving with each iteration, over many
iterations there is a clear oscillation. For systems of higher
SNR and/or decreased $\chi$ we find the MSD result to be very
close to BP and the theoretical result. The BP algorithms
reproduce the equilibrium result to within a small error after
only a few iterations, even in systems with only $600$ users and
$1000$ chips ($\chi=3/5$), across a range of $\gamma$. Where
unique saddle-point solutions were predicted by the equilibrium
analysis decoding by BP normally produced a stable fixed point.
The MSD results are not shown in subsequent figures but are
suboptimal with respect to BP in all cases.

A histogram of achieved BERs is demonstrated in figure
\ref{normal_decoding2}. In the large system limit we would expect
the cumulative distribution functions to converge towards step
functions (self averaging) on the one of the thermodynamic
solutions, but it is clear we are quite far from this scenario. BP
converges quickly towards results of very low or zero BER. The MSD
algorithm works very well, but more slowly than BP, for a subset
of examples. In many other samples the performance deteriorates as
 MSD is iterated beyond the initial approximation (matched filter).

If we consider a similarly sized system of $600$ chips and $1000$
users ($\chi=5/3$) then we can observe decoding performance in a
regime which should be characterized by metastability in the dense
code but not in the sparse code. The final $\BER$ achieved in
$300$ samples for various systems is shown as a cumulative
probability distribution in figure \ref{decodingbadly} after 10
iterations and after 80 iterations. The sparse system is unimodal,
with fast convergence in most systems. The dense ensemble is
multimodal as expected, the convergence time towards the low
$\BER$ solutions are very slow, and the majority of achieved
solutions are close to the high $\BER$ metastable solution. Random
initial conditions tend to produce steady states characterized by
the bad solution, even if this corresponds to the metastable
rather than equilibrium solution. The composite system equilibrium
solution is unique for $\gamma=8/9$. For $\gamma=8/9$ some $40\%$
of samples improve between iteration $10$ and iteration $80$, but
$40\%$ also worsen, the median performance is quite far from the
equilibrium value. The equilibrium results for $\gamma=0.5$ are
not closely approximated in the decoding experiments, the
performance in $\BER$ is worse everywhere than the equilibrium
prediction and the constituent dense and sparse systems. For large
$\chi$ it appears the finite size effects are more limiting in the
case of the composite codes, particularly at intermediate values
of $\gamma$.

The composite systems shown in figure \ref{decodingbadly} does not
come close to the performance of even the bad solution in either
the median or mean for this system size except for large or small
$\gamma$. The ability of the composite algorithm
~\cite{Mallard:BPDG} is limited in achieving the equilibrium
result for intermediate $\gamma$ than standard methods for the
sparse~\cite{Montanari:BPB} and dense~\cite{Kabashima:SMA}
ensembles for systems of this size. A quantitative comparison of
the equilibrium and finite size systems in the metastable regime
with bulk statistics such as the mean is difficult due to the
multimodal nature of the distributions.

The decoder performance for systems of size $O(1000)$ seem to
provide mean values for the BER, which are quite far from the
theoretical values and unable to realise the advantages of some
composite codes predicted by the equilibrium analysis. There are
many approximations made at $O(1/N)$ in construction of the BP
algorithm~\cite{Kabashima:SMA}~\cite{Mallard:BPDG}, which are both
systematic and random, and it is possible these are at the root of
the BP instability for intermediate values of $\gamma$. However, it is noteworthy that even without elimination of dense messages, performance with the proposed update schemes are poor at intermediate $\gamma$. The ordering of updates may also be important, and other sensible schemes might be consider. For example to iterate only the sparse messages until convergence (a fast process) between updates of dense message dependent quantities.

\subsection{Regular sparse ensembles}
\label{sec.regularensembleresults}
\begin{figure}[!htbp]
\includegraphics[width=0.6\linewidth]{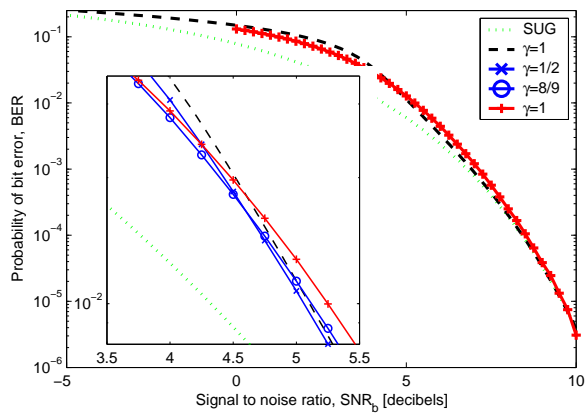}
\caption{(colour online) \label{fig:EnsembleComparison} Shown is the optimal
performance for $\gamma=\{0,1/2,8/9,1\}$,$\chi=1$, with a chip regular
sparse ensemble as a component in the composite system. At high
SNR the performance decreases with $\gamma$, at low SNR the
performance increases with $\gamma$. For a small range of
inclusive of the inset range, the composite codes outperform both
the sparse and dense codes.}
\end{figure}

A random ensemble with regular chip connectivity~\cite{Raymond:SS}
as well as regular user connectivity will be examined. There is
not a strong case for use of these codes in practical scenarios,
but they offer the possibility to demonstrate a positive result
for the composite system which is reproducible in small system
size experiments.

An interesting alternative composite system to the user regular
ensemble is one in which the number of accesses per chip (call
this $L$) is equal for all chips. This requires global
coordination of all users, with the loosening of this restriction
neither the doubly-regular ensemble presented nor the random codes
would represent the most high performing options. However, the
regular code is interesting because it shares many of the
topological features of the sparse random ensemble, but has very
slightly lower MAI, reduced by a factor $(L-1)/L$
~\cite{Raymond:RM} over the random sparse or dense ensemble, and
fewer finite size effects in the BP algorithm. The reduced MAI has
the effect that at low values for SNR the unique stable
solutions for the doubly regular ensemble is superior in $\BER$ to
the dense ensemble. With this ensemble one can demonstrate  a
statistically significant result, in decoding by BP, for which the
composite code ensemble outperforms the corresponding sparse and
dense parts in $\BER$.

The equilibrium behaviour of the doubly regular sparse ensemble
was analysed in~\cite{Raymond:SS}. Replacing the standard sparse
code ensemble by the composite one we can repeat the replica
analysis and use the same set of BP equations (identical at
leading order). In the analysis we find that the composite code
performance interpolates the sparse and dense performance in low
and high SNR regimes. However, in an intermediate range of
SNR, where the bit error rate is equal in the sparse and dense
models, the unique solutions for the composite ensembles become
superior to both the dense and sparse ensemble result. The
performance of several ensembles is shown in
figure~\ref{fig:EnsembleComparison}.

\begin{figure}[!htbp]
\includegraphics[width=0.8\linewidth]{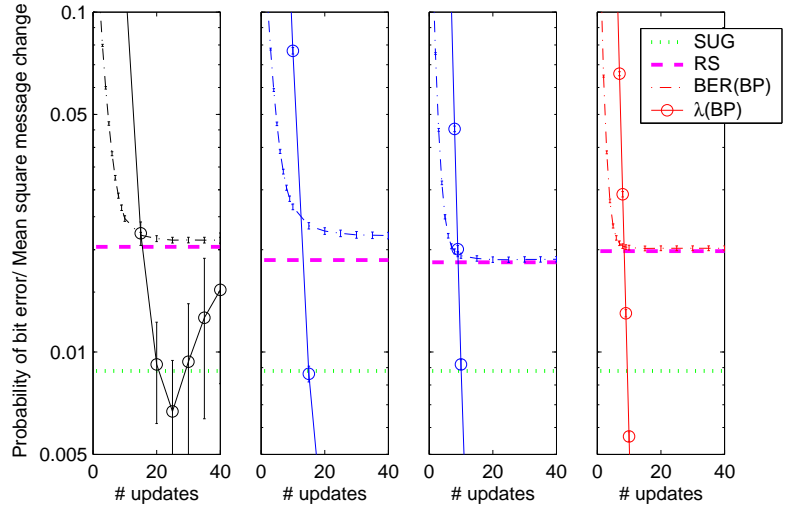}
\caption{(colour online) \label{fig:finallysuccess} At $\SNRb=4.5$,$\chi=1$, we show the mean results of $500$ decoding experiments using $\gamma=\{0,1/2,8/9,1\}$ (from left to right) with a chip regular sparse component in each composite system.
The BP equations converge except $\gamma=0$, where some samples
were unstable. A similar effect is manifested in the
$\gamma=\frac{1}{2}$ ensemble after about 45 iterations, but not
within the scale of the figure. Each set of samples produced a
$\BER$ in decoding close to the RS prediction except for
$\gamma=1/2$, where decoding performance was substantially poorer.
The $\BER$ by the RS result and sampling is best amongst ensembles
with $\gamma\approx 8/9$.}
\end{figure}

Working with a simulation of $1000$ users and $1000$ chips we are
able to demonstrate that the mean performance of several composite
codes exceed the performance of there constituent parts, as shown
in figure \ref{fig:finallysuccess}. The results for
$\gamma=\{0,8/9,1\}$ ensembles are close to the large system limit
prediction, to within error bars. The composite code achieves the
lowest bit error rate in expectation amongst the codes, and has
convergence properties somewhere between the two extremal
ensembles. As in previous experiments the performance for
$\gamma=1/2$ is much poorer than the large system limit
prediction.

As can be
seen the dense code fields are initially converging in a similar
way to figure \ref{normal_decoding}. However, at later time the
fields appear to be becoming unstable again (at least for a subset
of simulations that dominate the measure). This instability might be a direct consequence of the inaccuracy
of the Gaussian BP approximation (\ref{Zmuk}) when $\BER$ in
decoding becomes very small. Similar trends are seen in some of
the composite codes, often the messages do not converge exactly
but only to with some fixed variability.

\section{Discussion}

The equilibrium analysis demonstrates that in regions of
metastability the composite coding structure, comprising of a
sparse and densely connected component, might have some
interesting and valuable properties. When power is approximately
equal in the two parts performance is very close to the dense
ensemble, but with only a small amount of power in the dense code
properties are strongly distinguishable. At the same time, we have
shown that in reasonably sized simulations our algorithmic
approaches based on O(1/N) approximations in the dense code prove
to work relatively poorly in the composite codes. This instability
in some composite codes can persist even in scenarios where the
equilibrium analysis predicts a regime without a metastable
solution, which would be the natural candidate for a
non-equilibrium attractor.

We predict the failure of the composite BP algorithm to be in part
related to the Gaussian approximation in marginalization over
states (\ref{Zmuk}), which may be a poor approximation when
messages become strongly biased. If this is the case then the
problem may be avoided or mitigated by standard algorithmic tricks
such as annealing or damping. When the messages become very
biased, replacing the full marginalization by one considering only
a truncated set of states might be a viable polynomial time
alternative to using the analytical Gaussian approach. In small
realisations of composite systems many heuristics might be
employed.

In the final section we presented results with a different type of
sparse ensemble in which there was some coordination between
users. Similarly there would be some value in considering either
an ordered (optimised) sparse code combined with a random dense
code or vice-versa. The ordered sparse code might provide a method
for decoding under ideal channel conditions, whereas the dense
code provides a contingency and some of the advantages of the
spread spectrum approach. At least in so far as a practical method
this might be a well-motivated scenario.

The composite code presents an interesting dichotomy in its
surpression of the metastable behaviour, but greater apparent
instability in simulation. Aside from standard convergence
measures in simulations we have not thought of a concrete way to
probe the origins of this instability in simulations. One
possibility would be to study the equilibrium properties of the
metastable state in more detail in the thermodynamic framework.

A finite size scaling of the algorithm results would be valuable,
unfortunately the need to manipulate an $N$ by $K$ dense spreading
matrix restricts us somewhat in moving to larger scales. The
scales we have presented, and error bounds, are presented in a
somewhat ad-hoc way but this is only to demonstrate the breadth of
behavior. Many results in the cited papers go much further in
dealing with the question of finite size effects in sparse and
dense systems.

We have developed a method which is applicable where the sparse
ensemble includes graphs which are fully connected. If we consider
a composite code with a sparse part below the percolation
threshold a more fruitful analysis might consist in a two level
analysis. By considering the different classes of disconnected
components (within the sparse code) as the microscopic states
connected through a homogeneous (dense code) interaction. This may
also be the basis for better algorithms.

\section*{Acknowledgements}
Support from EPSRC grant EP/E049516/1 is gratefully acknowledged (DS).

\UPDATEBIBNO{

}
\UPDATEBIB{
\bibliographystyle{unsrt}
\bibliography{BibliographyMAY09}
}
\end{document}